\documentclass{article}

\newcommand{\eqref}[1]{(\ref{#1})}

\begin{document}

\title[Macroscopic Einstein Equations for a Cosmological Model]{Macroscopic Einstein Equations for a Cosmological Model with $\lambda$ - Term}
\author{Yurii Ignat'ev\\
Lobachevsky Institute of Mathematics and Mechanics, Kazan Federal University,\\ Kremleovskaya str., 35, Kazan 420008, Russia}

\begin{abstract}
 Through averaging the Einstein equations over transverse gravitational perturbations it is obtained a closed system of two ordinary differential equations describing macroscopic cosmological evolution of the isotropic space-flat Universe filled with gravitational radiation. It is found an asymptotic solution of evolution equation for gravitational perturbation amplitude. Making the substitution of this solution into Einstein equation averaged over gravitational perturbations, the single evolution non-linear ordinary differential second-order equation relative to macroscopic scale factor is obtained. It is also found a solution of evolution equation for scale factor in WKB-approximation which analytically describes the process of transformation from ultrarelativistic regime of cosmological extension to inflationary one.
%%%%%%%%%%%%%%%%%%%%%%
\end{abstract}

\section{Gravitational Perturbations of Isotropic Universe}
Gravitational perturbations of homogenous isotropic Universe were first introduced and classified  in pioneer work by E.M. Lifshitz, 1946 \cite{Lifshitz} (see e.g. the well-known L.D. Landau and E.M.Lifshitz monograph \cite{Land_Field}).
In classical works of R.A. Isaakson, 1968 \cite{Isaakson1,Isaakson2} there was developed the approach to build macroscopic gravitation theory based on averaging of microscopic (short-wave) fluctuation of metrics. In Author's works 1985-1991
(see e.g. \cite{Bogolyub}, 2007) it was being developed the statistical theory of gravitational interaction based on combination of ideas expressed in R.A. Isaakson's and N.N. Bogolyubov's works. In particular, these ideas were realized in \cite{Yu_astr} for dynamical derivation of a kinetic equation for photons against a background of locally fluctuating but macroscopically homogenous isotropic Universe. In the article we realize these ideas for a case of macroscopically flat Universe described by Einstein equations with a cosmological term assuming that single matter type in this Universe is presented by gravitational waves i.e. transverse traceless gravitational perturbations. As is well known, in absence of these perturbations the Universe is described by de Sitter solution or, if using synchronous system of coordinates - by inflationary solution.
Therefore not only process of derivation of closed evolution equations describing macroscopic Universe is of interest for us but also the research of such dynamically reasoned macroscopical model's behavior.

\subsection{Averaging of Einstein Equations}
According to general approach to derivation procedure of macroscopic equations \cite{Bogolyub} let us represent space-time metrics in form:
\begin{equation}\label{g+h}
g_{ik}=g^{(0)}_{ik}+\delta g_{ik}, \quad \delta g_{ik}=\epsilon g_{ik}
\end{equation}
where $\epsilon\ll 1$  è $g^{(0)}_{ik}$ is a macroscopic metric of space-time obtained using certain averaging:
\begin{equation}\label{average_g}
g^{(0)}_{ik}\equiv \overline{g_{ik}},
\end{equation}
so that:
\begin{equation}\label{average_dg}
\overline{\delta g_{ik}}\equiv 0.
\end{equation}
Assuming averaging operation to be independent on coordinates, let us also require derivatives of all arbitrary perturbations averages to be equal to zero:
\begin{equation}\label{average_ddg}
\overline{\partial_j \delta g_{ik}}=0;\quad \overline{\partial_{jl} \delta g_{ik}}=0.
\end{equation}
Let us consider Einstein equations with cosmological term\footnote{In the article Ricci tensor is defined through contraction of curvature tensor over first and third indices, metrics signature is $(---+)$.}:
\begin{equation}\label{Einst_lambda}
G_{ik}-\lambda g_{ik}=0,
\end{equation}
where $G_{ik}=R_{ik}-1/2Rg_{ik}$ is Einstein tensor.

Let us write Einstein equations with cosmological term in the second approximation over gravitational perturbations:
\begin{equation}\label{Einst0-2}
G^{(0)}_{ik}+G^{(1)}_{ik}+G^{(2)}_{ik}=\lambda g_{ik},
\end{equation}
where $G^{(0)}_{ik}=G_{ik}g^{(0)}_{ik}$, $G^{(1)}_{ik}=\mathrm{Lin}[G_{ik}(\delta g_{ik})]\sim \epsilon G^{(0)}_{ik}$,
$G^{(2)}_{ik}=\mathrm{Lin}[G_{ik}(\delta^2 g_{ik})]\sim \epsilon^2 G^{(0)}_{ik}$.
and average these equations taking into account relations (\ref{average_dg}), (\ref{average_ddg}):
Thus we get macroscopic second-order Einstein equations over perturbations of gravitational field:
\begin{equation}\label{Eq_<Einst2>}
G^{(0)}_{ik}=-\overline{G^{(2)}_{ik}}+\lambda g^{(0)}_{ik},
\end{equation}
according to which second-order corrections can be considered as gravitational perturbations' energy-momentum tensor\footnote{This statement is one of the main statements of R.A. Isaakson's theory \cite{Isaakson1,Isaakson2}}:
\begin{equation}\label{MET_gw}
T_{ik}=-\frac{1}{8\pi}\overline{G^{(2)}_{ik}}.
\end{equation}

\subsection{Transverse Perturbations of Space-Flat Isotropic Universe}
Let us write a metrics with monochromatic transverse gravitational perturbations of space-flat isotropic Universe in form (see e.g. \cite{Land_Field}):
\begin{eqnarray}
\label{Freed}
ds^2_0=a^2(\eta)(d\eta^2-dx^2-dy^2-dz^2);\\
\label{metric_pert}
ds^2=ds^2_0+a^2(\eta)h_{\alpha\beta}dx^\alpha dx^\beta;\\
\label{h(eta)}
h_{\alpha\beta}=e_{\alpha\beta}S(\eta)\mathrm{e}^{i\mathbf{nr}},
\end{eqnarray}
where $S(\eta)$ is an amplitude of gravitational waves. Thus, it is:
\begin{eqnarray}\label{Freed_g0}
g^{(0)}_{ik}=a^2(\eta)\mathrm{Diag}(-1,-1,-1,+1);\\
\delta g_{4\alpha}=0;\quad \delta g_{\alpha\beta}=a^2(\eta)h_{\alpha\beta}.
\end{eqnarray}
Then:
\begin{eqnarray}\label{defh1}
 h^\alpha_\beta=h_{\gamma\beta}g^{\alpha\gamma}_0\equiv-\frac{1}{a^2}h_{\alpha\beta};\\
 \label{defh2}
h\equiv h^\alpha_\alpha\equiv  g^{\alpha\beta}_0h_{\alpha\beta}\equiv
 -\frac{1}{a^2}(h_{11}+h_{22}+h_{33}) ,
\end{eqnarray}
herewith for transverse perturbations it is:
\begin{eqnarray}
\label{perp}
h^\alpha_\beta n_\alpha=0;\\
\label{perp1}
h=0.
\end{eqnarray}
As a result of (\ref{perp1}) in approximation linear over $h$ it is:
\begin{equation}\label{g}
\sqrt{-g}\approx \sqrt{-g_0}=a^4.
\end{equation}

by all directions of wave vector $\mathbf{n}$\footnote{It is also possible to carry out averaging over all lengths of wave vectors but this operation does not provide additional information. See  \cite{Bogolyub,Yubook1} about averaging procedure and getting Einstein macroscopic equations
.}

\subsection{Averaging Perturbations over Wave Vector Direction}
 Thus, we expand Einstein tensor into series in smallness of gravitational waves $S(\eta)$. Herewith using isotropy of unperturbed metrics it is convenient to introduce local system of coordinates where :
\begin{equation}\label{loc_coord}
\mathbf{n}=n(0,0,1); \quad \mathbf{s}=(1,0,0),
\end{equation}
where $\mathbf{s}$ is a unit vector of transverse perpetrations' polarization. In this system of coordinates it is
\begin{equation}\label{nz}
h_{12}=0; \ h_{11}=-h_{22}=S(\eta)\mathrm{e}^{inz},
\end{equation}
In arbitrary Cartesian coordinate system of 3-dimensional Euclid space $E_3$ polarization tensor $e_{ik}$ of formula (\ref{defh2}) has a form:
\begin{eqnarray}\label{e_ab}
e_{\alpha\beta}=2s_\alpha s_\beta+\frac{n_\alpha n_\beta}{n^2}-\delta_{\alpha\beta},\\
\label{orto3}
\mathbf{s}^2=1; \quad \mathbf{sn}=0,\quad \mathbf{n}^2=n^2.
\end{eqnarray}
It is easily checked that calibration condition (\ref{perp}) is automatically fulfilled in such a case.

Let us note that against a background of isotropic space, operation of averaging over directions is reduced to calculating an integral over 2-dimensional sphere of radius $n$:
\begin{equation}\label{average}
\overline{\phi(n,\mathbf{r})}=\frac{1}{4\pi}\int \phi(\mathbf{n},\mathbf{r})d\Omega_n.
\end{equation}
Thus, according to (\ref{average_dg}), (\ref{average_ddg}) we have:
\begin{eqnarray}
\overline{h_{\alpha\beta}} =0; \quad \overline{n_\gamma h_{\alpha\beta}} =0; \ldots
\end{eqnarray}

\subsection{Zero Approximation}
Expanding Einstein tensor in metrics perturbation, we get known expressions in zero approximation:
\begin{eqnarray}\label{Einst0}
G^{(0)}_{11}=G^{(0)}_{22}=G^{(0)}_{33}=2\frac{a''}{a}-\frac{a'^2}{a^2};\\
G^{(0)}_{44}=3\frac{a'^2}{a^2}.
\end{eqnarray}
Thus, in zero over gravitational perturbations approximation, we could get standard Einstein equation with $\lambda$ - term
\begin{equation}\label{EqEinst0}
\frac{a''}{a^4}=\frac{1}{3}\lambda
\end{equation}
and its inflationary solution:
\begin{equation}\label{inflat0}
a=-\frac{1}{\eta}\sqrt{\frac{3}{\lambda}}.
\end{equation}
%

%%%%%%%%%%%%%%%%%%%%%%%
\subsection{Equation for Gravitational Wave Amplitude}

In approximation linear over $S$ we get single non-trivial components:
\begin{eqnarray}\label{dG1}
G^{(1)}_{11}=-G^{(1)}_{22}\equiv \delta G= \frac{1}{2}\mathrm{e}^{i\mathbf{nz}}
\biggl[S''+2\frac{a'}{a}S'+S\biggl(n^2+2\frac{a'^2}{a^2}-4\frac{a''}{a}\biggr)\biggr].
\end{eqnarray}
Covariantly generalizing the result in $E_3$, let us write:
\begin{equation}\label{dG1_cov}
G^{(1)}_{\alpha\beta}=(\delta_{\alpha\beta}-2s_\alpha s_\beta)\delta G.
\end{equation}
Substituting expression (\ref{dG1}) in Einstein equations (\ref{Einst_lambda}), we get equation for an amplitude of transverse perturbations:
\begin{equation}\label{Eq_h}
S''+2\frac{a'}{a}S'+S\biggl(n^2+2\frac{a'^2}{a^2}-4\frac{a''}{a}+\lambda a^2\biggr)=0.
\end{equation}
This equation with an account of relations (\ref{Einst0}) can be written in simplified form:
\begin{equation}\label{GW_Eq}
S''+\frac{2}{\eta}S'+(n^2-2G^{(0)}_{11}+\lambda a^2)S=0.
\end{equation}
In particular, substituting zero approximation inflationary solution of Einstein equations (\ref{inflat0}) here, equation (\ref{GW_Eq}) can be reduced to the next one:
\begin{equation}\label{GW_Eq_0}
S'' -2\frac{S'}{\eta}+S\left(n^2-\frac{3}{\lambda\eta^2}\right)=0,
\end{equation}
which has its solution:
\begin{equation}\label{Sol_h_inflat}
S=C_1 \eta^{3/2}\mathrm{J}_\mu(n\eta)+C_1 \eta^{3/2}\mathrm{Y}_\mu(n\eta),
\end{equation}
\begin{equation}\label{mu}
\mu=\frac{\sqrt{3}}{2}\sqrt{3+4/\lambda}.
\end{equation}
Particularly, near cosmological singularity of zero approximation $\eta\to-\infty$, hence $|\mathbf{n}\eta|\to \infty$
equation (\ref{GW_Eq}) is reduced to the next oscillation equation:
\begin{equation}\label{neta8}
S''+n^2S=0,
\end{equation}
having its solution ordinary WKB-solutions:
\begin{equation}\label{WKB_Sol}
S=C_1e^{in\eta}+C_2e^{-in\eta},
\end{equation}
which can be obtained also from the exact solution of (\ref{Sol_h_inflat}) in this limit.

\subsection{Second Approximation}
Calculating Einstein tensor of second approximation we obtain its non-trivial components:
\begin{eqnarray}
\label{G2_11}
G^{(2)}_{11}=G^{(2)}_{22}=\mathrm{e}^{2inz}\biggl(\frac{5}{4}S^2n^2+SS''+\frac{1}{4}S'^2+2\frac{a'}{a}SS'  \biggr),\\
\label{G2_33}
G^{(2)}_{33}=\mathrm{e}^{2inz}\biggl(\frac{1}{4}S^2n^2+SS''+\frac{3}{4}S'^2+2\frac{a'}{a}SS'  \biggr),\\
\label{G2_44}
G^{(2)}_{44}=-\mathrm{e}^{2inz}\biggl(\frac{7}{4}S^2n^2+\frac{1}{4}S'^2+2\frac{a'}{a}SS'  \biggr).
\end{eqnarray}
Carrying out covariant generalization in $E_3$, we can write:
\begin{equation}\label{G_ab_cov}
G^{(2)}_{\alpha\beta}=\mathrm{e}^{2i\mathbf{nr}}\bigl(U\delta_{\alpha\beta}-V\frac{n_\alpha n_\beta}{n^2}\bigr),
\end{equation}
where
\begin{eqnarray}\label{U}
U=\frac{5}{4}S^2n^2+SS''+\frac{1}{4}S'^2+2\frac{a'}{a}SS'  ;\\
\label{V}
V=S^2n^2-\frac{1}{2}S'^2.
\end{eqnarray}
Averaging (\ref{G_ab_cov}) over directions of perturbation propagation with an account of obvious equality
\begin{equation}\label{<nn>}
\overline{n_\alpha n_\beta}\equiv \frac{1}{3}\delta_{\alpha\beta} n^2,
\end{equation}
we get for averaged components $G^{(2)}_{ik}$ the next expression:
\begin{equation}\label{<G2_ik>}
\overline{G^{(2)}_{ik}}=8\pi(\mathcal{E}+\mathcal{P})u_iu_k-8\pi\mathcal{P}g_{ik},
\end{equation}
where $u^i$ is a timelike vector of observer velocity and $\mathcal{E}$ and $\mathcal{P}$ are energy density and pressure of transverse gravitational perturbations:
\begin{eqnarray}
\label{Egw}
 \mathcal{E}= \frac{1}{8\pi a^2}\biggl(\frac{7}{4}S^2n^2+\frac{1}{4}S'^2-2\frac{a'}{a}SS'  \biggr)\\
\label{Pgw}
\mathcal{P}=-\frac{1}{8\pi a^2}\biggl(\frac{11}{12}S^2n^2+\frac{5}{12}S'^2+2SS'\frac{a'}{a}+SS'' \biggr)
\end{eqnarray}

In particular, for WKB-solution (\ref{WKB_Sol}) these formulas lead to effective ultrarelativistic state equation:
\begin{eqnarray}
\label{Egw_WKB}
\mathcal{E}\approx\frac{3}{16\pi a^2}S^2n^2;\\
\label{Pgw_WKB}
\mathcal{P}\approx\frac{1}{16\pi a^2}S^2n^2 =\frac{1}{3}\mathcal{E}.
\end{eqnarray}

\section{Macroscopic Einstein Equations of Second Order of Gravitational Perturbations for Isotropic Space-Flat Universe}
\subsection{Evolution Equations}
Combining obtained results within the frame of equations (\ref{Einst0-2}) and (\ref{Eq_<Einst2>}), we obtain a self-consistent system of ordinary second-order differential equations describing cosmological evolution of flat macroscopic Universe with an account of transverse gravitational perturbations:
\begin{eqnarray}
\label{evolut_S}
S''+2\frac{a'}{a}S'+S\biggl(n^2+2\frac{a'^2}{a^2}-4\frac{a''}{a}+\lambda a^2\biggr)=0;\\
\label{evolut_a}
3\frac{a'^2}{a^2}=\frac{7}{4}S^2n^2+\frac{1}{4}S'^2-2\frac{a'}{a}SS'+\lambda a^2.
\end{eqnarray}
In such a case equation (\ref{evolut_S}) describes cosmological evolution of scalar amplitude $S(\eta)$ of gravitational perturbations and equation (\ref{evolut_a}) describes cosmological evolution of scale factor $a(\eta)$. In this case evolution equation for gravitational perturbations is linear differential equation of second order relative to amplitude of these perturbations.

Let us notice that if there is a perturbation spectrum instead of monochromatic perturbations (\ref{h(eta)}):
\begin{equation}\label{spectr_h}
h_{\alpha\beta}=\frac{1}{(2\pi)^{3/2}}\int\limits e_{\alpha\beta}(\mathbf{n})\left[S_n(\eta)\mathrm{e}^{-i\mathbf{nr}}+
S^*_n(\eta)\mathrm{e}^{-i\mathbf{nr}}\right]d^3\mathbf{n},
\end{equation}
then in evolution equation (\ref{evolut_S}) it is necessary to make a substitution $S\to S_n$ and in evolution equation for scale factor (\ref{evolut_a}) it is necessary to use expressions for averages:
\begin{eqnarray}\label{average_S2_spectr}
n^2S^2\to\overline{n^2S^2}=\frac{1}{\sqrt{2\pi}}\int_0^\infty S_nS^*_n n^2 dn;\\
\label{average_S'2_spectr}
S'^2\to\overline{S'^2}=\frac{1}{\sqrt{2\pi}}\int_0^\infty S'_n{S^*_n}'dn;\\
\label{average_S'S_spectr}
S'S\to\overline{S'S}=\frac{1}{\sqrt{2\pi}}\int_0^\infty \frac{1}{2}(S'_nS^*_n+S_n{S^*_n}')dn.
\end{eqnarray}

\subsection{Asymptotic Solutio of Evolution Equation for Perturbations}

Let us notice that using scale transformation of amplitude $S$
\begin{equation}\label{trans_S}
S=\frac{\phi}{a}
\end{equation}
in equation (\ref{evolut_S}) it is possible to get rid of first derivative:
\begin{equation}\label{Eq_phi}
\phi''+Q(n,\eta)\phi =0,
\end{equation}
where
\begin{equation}\label{Q}
Q(n,\eta)=n^2+2\frac{a'^2}{a^2}-5\frac{a''}{a}+\lambda a^2.
\end{equation}
Let us consider asymptotics of solutions of this equation assuming
\begin{equation}\label{WKB}
n\gg 1; \quad Q(n,\eta)>0.
\end{equation}
Using theorem \cite{Fedoruk} about asymptotic solution of equation (\ref{Eq_phi}), let us write its asymptotic independent solutions:
\begin{equation}\label{asymptot}
\phi(\eta)\sim Q(n,\eta)^{-1/4}\exp\biggl\{\pm i\int\limits_{\eta_0}^{\eta} \sqrt{Q(n,\eta')}d\eta' \biggr\}
\equiv Q(n,\eta)^{-1/4}\mathrm{e}^{\pm i\Phi(n,\eta)} ,
\end{equation}
and asymptotic values of derivatives (taking into account equations (\ref{Q})):
\begin{eqnarray}\label{phi'}
\phi'(\eta)\sim \pm iQ(n,\eta)^{1/4}\exp\biggl\{\pm i\int\limits_{\eta_0}^{\eta} \sqrt{Q(n,\eta')}d\eta' \biggr\}
\equiv \pm iQ(n,\eta)^{1/4}\mathrm{e}^{\pm i\Phi(n,\eta)} ;\\
\phi''(\eta)\sim - Q(n,\eta)^{3/4}\exp\biggl\{\pm i\int\limits_{\eta_0}^{\eta} \sqrt{Q(n,\eta')}d\eta' \biggr\}\equiv
-Q(n,\eta)^{3/4}\mathrm{e}^{\pm i\Phi(n,\eta)} .
\end{eqnarray}

Further, for us it is important the fact that terms which are, practically, quadratic over amplitudes and included in the right part of evolution equation
(\ref{evolut_S}) are calculated as a square of complex variable's modulus i.e. $S^2=SS^*$, where $S$ and $S^*$ are two independent asymptotics (\ref{asymptot}). Therefore it is:
\begin{eqnarray}\label{S^2}
S^2\to SS^*\equiv \frac{\phi\phi*}{a^2}; \; S'^2\to S'{S^*} '=\nonumber\\
\frac{\phi'{\phi^*} '}{a^2}-
(\phi{\phi^*} '+\phi'\phi*)\frac{a'}{a^2}+\phi\phi^*\frac{a'^2}{a^4}\equiv \frac{\phi'{\phi^*} '}{a^2}
+\phi\phi^*\frac{a'^2}{a^4};\nonumber\\
SS'\to \frac{1}{2}(S'S^*  +S{S^*} ')=\frac{1}{2a^2}(\phi'\phi^*+\phi{\phi^*}')-\frac{a'}{a^3}\phi\phi^*\equiv -\frac{a'}{a^3}\phi\phi^*.
\end{eqnarray}

Calculating, we find:
\begin{eqnarray}\label{S2}
S^2\simeq \frac{S^2_0}{a^2(\eta)\sqrt{Q(n,\eta)}};\; SS'\simeq -\frac{S^2_0}{a^2(\eta)\sqrt{Q(n,\eta)}}\frac{a'}{a},\nonumber\\
S'^2\simeq S^2_0\frac{\sqrt{Q(n,\eta)}}{a^2}+S^2_0\frac{a'^2}{a^4(\eta)\sqrt{Q(n,\eta)}};\;
\end{eqnarray}
where $S_0$ is a constant.

Substituting these expressions in equation of scale factor evolution (\ref{evolut_a}), we finally get macroscopic evolution equation in asymptotic approximation:
\begin{eqnarray}
\label{evolut_a_asympt}
3\frac{a'^2}{a^4}-\lambda=\frac{S^2_0}{a^4}\biggl(
\frac{7}{4}\frac{n^2}{\sqrt{Q(n,\eta)}}+\frac{1}{4}\sqrt{Q(n,\eta)}+\frac{9}{4}\frac{a'^2}{a^2}\frac{1}{\sqrt{Q(n,\eta)}}\biggr).
\end{eqnarray}
Òhus, we obtained a closed ordinary highly non-linear differential second-order equation\footnote{let us remind that function $Q(n,\eta)$ depends on $a,a',a''$}, which provides an asymptotically exact description of macroscopic cosmological evolution of the Universe filled with weak transverse gravitational perturbations. Basically, this equation can be investigated by methods of qualitative theory of differential equations.
We intend to revisit this problem in next publications. Now let us consider WKB-approximation of equation (\ref{evolut_a_asympt}).

\subsection{WKB-Solution of Evolution Equation for Scale Factor}
Let us consider now the following WKB-approximation of evolution equations:
\begin{equation}\label{WKB-approx}
n\gg \frac{a'}{a}; \quad n^2\gg \frac{a''}{a}\Rightarrow n\eta\gg 1.
\end{equation}
In this approximation
\begin{equation}\label{Q_approx}
Q(n,\eta)\approx n,
\end{equation}
equation (\ref{evolut_a_asympt}) takes fairly simple form:
\begin{equation}\label{evolut_a_WKB}
3\frac{a'^2}{a^4}-\lambda=\frac{2S^2_0n}{a^4}
\end{equation}
Solution of this equation can be written in elementary functions, proceeding to physical time $t$, so that $ad\eta=dt$. Carrying out elementary integration, we find:
\begin{equation}\label{WKB-solve}
a(t)=\biggl(\frac{2S^2_0 n}{\lambda}\biggr)^{1/4}\sqrt{\sinh \biggl(2\sqrt{\frac{\lambda}{3}}t\biggr)}.
\end{equation}
Solution of such form was obtained by the Author (see e.g. \cite{Yu_trans}). This solution describes smooth transition from ultrarelativistic stage of Universe expansion to inflationary stage at $t>t_c=\sqrt{3/4\lambda}$.

Actually, at $\to0$ we get from (\ref{WKB-solve}):
\begin{equation}\label{t0}
A(t)\approx \biggl(\frac{8S^2_0 n}{3}\biggr)^{1/4}\sqrt{t},
\end{equation}
and at $t\to\infty$ we get from (\ref{WKB-solve}):
\begin{equation}\label{t8}
a(t)\approx \biggl(\frac{S^2_0 n}{4\lambda}\biggr)^{1/4}\mathrm{e}^{\sqrt{\lambda/3}t}.
\end{equation}

\section{Discussion of Results}
Summarizing the results of the article, let us indicate the most important ones:
\begin{enumerate}
\item Equations of second order over perturbation amplitude are obtained from Einstein equations with $\lambda$ - term using expansion of metrics in small transverse perturbations relative to background Friedmann solution.
\item Eventually, a closed mathematical model describing cosmological evolution of the Universe filled with gravitational radiation, is obtained. This model consists of two ordinary second-order differential equations which we call evolution equations for the sake of brevity.
\item First equation for evolution of amplitude of gravitational perturbations' monochromatic mode is a linear homogenous differential equation.
\item Second evolution equation describes evolution of macroscopic scale factor of the Friedmann Universe. This equation is highly non-linear and is defined trough solutions of evolution equation for amplitude of gravitational perturbations.
\item An asymptotic solution of evolution equation for amplitude of gravitational perturbations is found.
\item Using obtained solution there have been calculated the macroscopic averages of perturbation amplitude's square and its derivatives.
\item Eventually, closed highly non-linear ordinary differential equation describing cosmological evolution of scale factor is obtained. Energy spectrum of gravitational perturbations and cosmological constant are parameters of this equation.
\item It is found the WKB-solution of this evolution equation analytically describing transformation from ultrarelativstic stage of the Universe expansion to inflationary one.
\end{enumerate}
Thus, as was noticed in \cite{unstab}, an account of gravitational perturbations in energy balance of the Universe at its early stages really discards primordial inflationary expansion stage and puts it on the second place after ultrarelativstic stage.

Let us notice that in recent work
Chiu Man Ho and Stephen D.H. \cite{Chiu}, devoted to quantum instability of the de Sitter Universe due to particles generation it is obtained similar conclusion.

Further we will be purposing first, on investigation of solution of evolution equations by numerical methods and second, on construction of equivalent mathematical model accounting other types of gravitational perturbations and physical fields.

\section*{Acknowledgement}

This work was founded by the subsidy allocated to Kazan Federal University for the state assignment in the sphere of scientific activities.

In conclusion, Author expresses his gratitude to participants of MW seminar for relativistic kinetics and cosmology of Kazan Federal University for helpful discussion of the work.

%%%%%%%%%%%%%%%%%%%%%%%%%%%%%%%%%%%%%%%%%%%%%%%%%%%%%%%%%%%%%%%%%

%


\begin{thebibliography}{30}
%
\bibitem{Lifshitz}
Å.Ì. Lifshitz,  JETP, 1946, 16, 697.
%
\bibitem{Land_Field}
L.D. Landau, E.M. Lifshitz. The Classical Theory of Fields. Pergamon Press.
Oxford$\cdot$ New York$\cdot$ Toronto$\cdot$ Sydney$\cdot$ Paris$\cdot$ Frankfurt, 1971.
%
\bibitem{Isaakson1}
R.A. Isaakson, Phys. Rev., 1966, v. 166, 1263.
%
\bibitem{Isaakson2}
R.A. Isaakson, Phys. Rev., 1966, v. 166, 1272.
%
\bibitem{Bogolyub}
Yu.G. Ignat'ev, Grav. and Cosmol., 2007,  v. 13, pp. 59-81.
%
\bibitem{Yu_astr}
Yu.G. Ignat'ev, A.A. Popov, Astrophys. Space Sci., 1990, v. 163, pp.
153-174
%
\bibitem{Yubook1}%6
Yurii G. Ignatyev (Ignat'ev). Relativistic Kinetic Theory of Nonequilibrium Processes in Gravitational
Fields. Kazan, Foliant-Press, -- 2010;  http://rgs.vniims.ru/books/const.pdf.
%
\bibitem{Fedoruk}
M.V. Fedoruk, Ordinary differential equations, Moskow, Lan, 2003.
%
\bibitem{Yu_trans}
Yu. G. Ignat'ev, Russ. Phys. J., 2013, v. 56, 693.
DOI: 10.1007/s11182-013-0087-4
%
\bibitem{unstab}
Yurii Ignat'ev, arXiv:1508.05375v1 [gr-qc].
%
\bibitem{Chiu}
Chiu Man Ho and Stephen D. H. Hsu, arXiv:1501.00708v2 [hep-th].
\end{thebibliography}
\end{document}